\documentclass[aps]{revtex4}   

\usepackage{graphicx}

\begin{document}

\title{Mean Field Theoretical Structure of He and Be Isotopes}

\author{ S.J. Lee\footnote{Email: ssjlee@khu.ac.kr}, Y.W. Son, and J.W. Park}
                    
\address{Department of Physics and Institute of Natural Sciences,
                    Kyung Hee University, Suwon, KyungGiDo, Korea}

%\rightline{\today}

\begin{abstract}

The structures of He and Be even-even isotopes 
are investigated using an axially symmetric Hartree-Fock approach 
with a Skyrme-IIIls mean field potential.
In these simple HF calculations, He and Be isotopes appear to be prolate 
in their ground states and Be isotopes have oblate shape isomeric states. 
It is also shown that there exists a level crossing when the nuclear shape 
changes from the prolate state to the oblate state.
The single neutron levels of Be isotopes
exhibit a neutron magic number 6 instead of 8 
and show that the level inversion between 
1/2$^-$ and 1/2$^+$ levels occurs only for a largely deformed isotope.
Protons are bound stronger in the isotope with more neutrons 
while neutron levels are somewhat insensitive to the number of neutrons
and thus the nuclear size and also the neutron skin become 
larger as the neutron number increases.
In these simple calculations with Skyrme-IIIls interaction 
no system with a clear indication of neutron halo
was found among He and Be isotopes. 
Instead of it we have found $^8$He+2n, 2n+$^8$He+2n, and $^{16}$Be+2n 
like chain structures with clusters of two correlated neutrons.
It is also shown that $^8$He and $^{14}$Be in their ground states are below 
the neutron drip line in which all nucleons are bound with negative energy and 
that $^{16}$Be in its ground state is beyond the neutron drip line 
with two neutrons in positive energy levels.
\end{abstract}


\maketitle
     
%\narrowtext

Owing to the progress of experimental techniques, information concerning 
the ground state and the excited states of light unstable nuclei has 
rapidly increased.
Furthermore radioactive ion beam accelerator will produce various unstable 
nuclei far from $\beta$-stability line.
In light stable nuclei, it has already been known that clustering is one of 
the essential features of nuclear dynamics not only in excited states 
but also in ground states.
Thus we may expect cluster structures in light unstable nuclei.
Pioneering theoretical studies have suggested the development of cluster structures
with a $2\alpha$ core in Be and B isotopes \cite{ref1,ref2,ref3,ref4,ref5}.
The $\alpha$ chain structures of $^8$Be and $^{12}$C was investigated 
using a relativistic mean field (RMF) approximation \cite{alpha}.
In the case of $^{12}$Be, the existence of cluster states was suggested 
in experimental measurements of the excited states \cite{ref6} and in 
experiments of $^6$He+$^6$He and $^8$He+$^4$He breakup reactions \cite{ref7}.
The measured spin-parities of excited states indicate a rotational band with a large
moment of inertia. These states are candidates of states with cluster structure.
It is an interesting subject to investigate clustering aspects in Be isotopes.
The cluster structures of $^{12}$Be were studied with a potential model with He 
clusters \cite{ref8} and with an algebraic version of resonating group 
model \cite{ref9} using $\alpha$ particles as inert clusters.
To eliminate the model assumption of inert $\alpha$ cluster an antisymmetrized 
molecular dynamics (AMD) \cite{ref10} is used in study of Be isotopes.
However this study uses inert Gaussian wave packets to represent nucleons.
These studies treat $\alpha$ particles as inert clusters or nucleon as 
a fixed Gaussian packet, and thus miss any detailed structure of nuclei 
and any change of the internal structure of $\alpha$ particles in a nucleus.
For a study of the detailed nuclear structure (such as size, shape, 
quadrupole moment, single nucleon levels) of these unstable nuclei,
we need to use a self-consistent mean field approach in terms of nucleon itself.
Mean field approaches have been applied successfully even for a small system 
such as nuclear ground states of $^6$Be, $^7$Be and $^7$B \cite{beamdhf,skhf}
and the Hartree-Fock (HF) energy of atomic He ground state \cite{hehf}.
RMF calculation with a nonlinear scalar field estimated binding energy of $^4$He
moderately well \cite{alpha}.
The ground states of nuclei with $2 \le Z \le 114$ around proton drip line
have been studied using Skyrme HF+BCS \cite{hfbcs}. 
Skyrme HF approach had described moderately well not only the ground state of $^4$He 
but also the two- and four-particle excited states of $^4$He \cite{he4xhf}.

An iterative method of intensity interferometry technique
had been applied to dissociation of the three established
two-neutron halo systems, $^6$He, $^{11}$Li and $^{14}$Be, and the n-n correlation
functions and corresponding source sizes had thus been extracted \cite{halosys}.
Based on the transverse momentum distributions after break-up of $^{11}$Li 
into $^9$Li+n+n,
it was suggested that the basic structure of $^{11}$Li is a $^9$Li-core 
surrounded by an extended cloud of two correlated neutrons \cite{hansen,johan}
at a large distance from the core.
It was pointed out \cite{neutclus} that the extra stability provided by 
correlations between many neutrons surrounding a nuclear core may result in stable, 
exotic structures along or perhaps even beyond the neutron drip line.
The ground state and low-lying states of $^{12}$Be presented us with other subjects 
concerning the vanishing of a neutron magic number 8 \cite{ref10}. 
The vanishing of the neutron magic number 8 
was already known in a neighboring nucleus, $^{11}$Be.
The vanishing in $^{12}$Be was predicted in the early theoretical work \cite{magic6}
using shell model configuration mixing.
Contribution from $^{10}$Be$\otimes (sd)^2$ in the low-lying states of $^{12}$Be
were experimentally studied by $^{10}$Be(t,p)$^{12}$Be reaction \cite{fortun}.
To study exotic nuclear structure without using configuration mixing
we will use non-spherical axially symmetric mean field theory here.

In a nuclear mean field theory, a nuclear system is composed of
nucleons interacting strongly through a mean field potential.
While a nonrelativistic mean field approach uses a phenomenological
potentials as a function of nucleon density,
the mean field potential in a relativistic treatment
is determined as a resultant of the meson exchange.
After Walecka proposed a relativistic mean field (RMF) approach
for a nuclear system in which Dirac nucleons interact
by exchanging classical meson fields \cite{walecka},
the theory has been extended to include various quantum effects and
successfully applied in describing nuclear matter and finite nuclei 
\cite{sjlee,deform,deform2,jhlee,sjlqr}.
However the usage of the relativistic mean field approach is mostly limited to
spherical or moderately deformed nuclei
since the RMF failed in describing highly deformed nuclei
unless using a nonlinear meson field \cite{sjlee,deform,deform2}.
On the other hand the nonrelativistic Hartree-Fock (HF) approach using empirical
Skyrme interaction can describe highly deformed nuclei moderately 
well \cite{zamick,tdhf} as well as describing spherical nuclei.
Thus we use nonrelativistic HF approach here in describing structures 
of He and Be isotopes.
As a first step, a simple HF method is used in this paper to investigate
various properties of the lowest energy states of He and Be even-even isotopes
before studying various excited states.

In a nonrelativistic HF approach, nucleon-nucleon nonlocal interactions are 
replaced by a density dependent Skyrme mean field potential which is a 
function of nuclear density and charge density in the nucleus. 
The level of each nucleon in a nucleus is determined 
by Schr\"{o}dinger equation in the mean field potential. 
By occupying nucleons in these single nucleon levels, 
we determine the nuclear and charge densities in the nucleus which
are then used to determine the new Skyrme mean field potential.
Then this new potential is used to determine new levels of nucleon.
We repeat this procedure until we get the converged results.
By occupying $Z$ single proton levels and $N$ single neutron levels 
starting from the lowest level, we can obtain the lowest energy state 
(either ground state or an excited local minimum in shape variation) of 
a nucleus with $Z$ protons and $N$ neutrons.
To discriminate from the particle-hole type excitation,
we will call the excited local minimum state by shape isomeric state here 
since the state would live longer than the particle-hole type excited state.
The numerical method and the Skyrme-IIIls interaction used here are exactly 
the same as that used in Ref.\cite{tdhf} which is applicable to a nuclear 
system with axial symmetry.
The axially symmetric two dimensional space is denoted here by the body fixed
symmetry axis $z$ and the radius $r$ from the symmetry $z$-axis.
The level of each nucleon is identified by $(\Omega,\pi_z,\pi_r)$
where $\Omega$ is the total magnetic quantum number of the single nucleon 
level in the intrinsic $z$-direction 
and $\pi_z$ and $\pi_r$ are the $z$- and $r$-parity in this $z$-$r$ space.
While $\Omega$ and $\pi_r$ stay to be a good quantum number during the
iterative calculation due to the axial symmetry, 
$\pi_z$ does not need to remain as a good quantum number unless we further 
restrict the system to have a reflection symmetry in $z$-direction too.

%%%  Table 1

%%%  Table 2

%%%  Table 3

%%%  Table 4

\begin{table} 
\caption{The binding energy per nucleon $E_B$ (MeV) with CM correction \cite{he4xhf},
the root mean square radius $R$ (fm), and the intrinsic quadrupole moment $Q$ 
(fm$^2$) of the lowest state of He isotopes.} \label{table1} 
\begin{tabular}{lccccccccccccccc}
\hline
   & \ \ \ $^4$He \ \ \ & \ \ \ $^6$He \ \ \ & \ \ \ $^8$He \ \ \ & \ \ \ $^{10}$He \ \ \
   & \ \ \ $^{12}$He \ \ \  \\
\hline
 $E_B$                
   &   7.15    &    5.13     &     4.30     &     3.04     &     2.27   \\
 $E_B$(exp) \cite{bedata}
   &   7.07    &    4.88     &     3.93     &     3.03     &    \\
 $R_{\rm proton}$      
   &   1.99    &    1.99     &     1.95     &     1.98     &     1.96   \\
 $R_{\rm neutron}$     
   &   1.98    &    2.79     &     2.86     &     8.12     &     9.92   \\
 $R_{\rm total}$        
   &   1.98    &    2.55     &     2.67     &     7.32     &     9.09   \\ 
 $Q_{\rm proton}$  
   &   0.03    &    0.72     &     0.13     &     0.25     &     0.11   \\
 $Q_{\rm neutron}$ 
   &   0.03    &    5.56     &     0.68     &     223.11   &   439.88   \\
 $Q_{\rm total}$   
   &   0.06    &    6.28     &     0.81     &     223.36   &   439.99   \\
\hline
\end{tabular}
\end{table}

\begin{table}
\caption{The level energy in MeV of single nucleon in level 
($\Omega$, $\pi_z$, $\pi_r$) of He isotopes in their ground state.
Each level is twofold degenerated with $\pm\Omega$.} \label{table2} 
\begin{tabular}{lccccccccccccccc}
\hline
 & $(\Omega,\pi_z,\pi_r)$ & \ \ \ $^4$He \ \ \ & \ \ \ $^6$He \ \ \ & \ \ \ $^8$He \ \ \ 
   & \ \ \ $^{10}$He \ \ \ & \ \ \ $^{12}$He \ \ \  \\
\hline
 proton & (0.5,+,+)  
   & --14.41   &  --22.99  & --31.51   &  --31.54   &  --31.91   \\
        & (0.5,--,+) 
   &    0.60   &   --9.70  & --14.85   &  --14.89   &  --15.20   \\
        & (1.5,+,--) 
   &    0.54   &   --4.57  & --13.86   &  --13.89   &  --14.43   \\
 neutron& (0.5,+,+)  
   & --15.37   &  --16.28  & --17.98   &  --17.96   &  --17.96   \\
        & (0.5,--,+) 
   &  --0.18   &   --1.96  &  --2.78   &   --2.80   &   --2.84   \\
        & (1.5,+,--) 
   &  --0.23   &   --0.84  &  --2.28   &   --2.30   &   --2.39   \\
        & (0.5,+,+)  
   &    1.62   &     0.77  &    0.77   &     0.71   &     0.67   \\
        & (0.5,--,+) 
   &    1.93   &     0.86  &    0.88   &     0.78   &     0.74   \\
\hline
\end{tabular}
\end{table}

%%%%%%    Be  %%%

\begin{center}
\begin{table} 
\caption{The binding energy per nucleon $E_B$ (MeV) with CM correction \cite{he4xhf},
the root mean square radius $R$ (fm), and the intrinsic quadrupole moment $Q$ 
(fm$^2$) of the ground state and the shape isomeric excited state of Be isotopes.} 
    \label{table3} 
\begin{tabular}{lccccccccccccccc}
\hline
   &  \ \ $^8$Be \ \  & \  \ $^8$Be \ \  & \  \ $^{10}$Be \  \ & \  \ $^{10}$Be \  \ 
   &  \ \ $^{12}$Be  \ \ &  \ \ $^{12}$Be  \ \ & \  \ $^{14}$Be \  \ 
   &  \ \ $^{16}$Be  \ \ &  \ \ $^{16}$Be  \ \ & \  \ $^{18}$Be \  \  \\
\hline
 $E_B$                
   &   6.71    &    6.04     &    6.72     &    6.51     &     6.04     &   6.03 
   &   5.25    &    4.49     &    4.25     &    3.83  \\
 $E_B$(exp) \cite{bedata}
   &   7.06    &             &    6.50     &             &     5.72     &  
   &   4.99    &    4.27     &             &     \\
 $R_{\rm proton}$      
   &   2.55    &    2.40     &    2.38     &    2.33     &     2.35     &   2.34  
   &   2.50    &    2.56     &    2.41     &    2.56  \\
 $R_{\rm neutron}$     
   &   2.53    &    2.38     &    2.55     &    2.51     &     2.76     &   2.75  
   &   3.17    &    3.42     &    3.97     &    7.01  \\
 $R_{\rm total}$        
   &   2.54    &    2.39     &    2.48     &    2.44     &     2.63     &   2.62  
   &   3.00    &    3.23     &    3.65     &    6.30  \\
 $Q_{\rm proton}$  
   &   7.17    &  --2.65     &    4.41     &  --2.24     &     2.50     & --1.75  
   &   5.93    &    6.64     &  --2.63     &    6.67  \\
 $Q_{\rm neutron}$ 
   &   7.07    &  --2.60     &    4.45     &  --2.86     &     1.66     & --1.05  
   &  16.97    &   23.44     &  --9.70     &  287.76  \\
 $Q_{\rm total}$   
   &  14.25    &  --5.25     &    8.86     &  --5.10     &     4.16     & --2.80  
   &  22.90    &   30.08     & --12.33     &  294.43  \\
\hline
\end{tabular}
\end{table}
\end{center}

\begin{center} 
\begin{table} 
\caption{The level energy in MeV of single nucleon in level 
($\Omega$, $\pi_z$, $\pi_r$) of Be isotopes in their ground state 
and shape isomeric state.
Each level is twofold degenerated with $\pm\Omega$.
Note here that the proton level $(1.5,+,-)$ is occupied instead of the 
lower level $(0.5,-,+)$ 
in the oblate $^{12}$Be.}
 \label{table4} 
%     \hspace*{-3.3cm}
\begin{tabular}{lccccccccccccccc}  
\hline
 & $(\Omega,\pi_z,\pi_r)$ &  $^8$Be  &  $^8$Be  &  $^{10}$Be  &  $^{10}$Be  
   &  $^{12}$Be  &  $^{12}$Be  &  $^{14}$Be  
   &  $^{16}$Be  &  $^{16}$Be  &  $^{18}$Be   \\
\hline
 proton & (0.5,+,+)  
   & --21.41   &  --23.49  &  --30.29  &  --31.01  & --35.87   &  --36.21   
   & --38.22   &  --41.23  &  --40.55  &  --41.34   \\
        & (0.5,--,+) 
   & --10.76   &   --6.96  &  --15.32  &  --14.05  & --19.57   & --20.14 
   & --24.95   &  --29.41  &  --24.48  &  --29.53   \\
        & (1.5,+,--) 
   &  --3.53   &   --7.87  &  --12.71  &  --14.48  & --19.55   & --19.26 
   & --20.85   &  --23.68  &  --25.28  &  --23.80   \\
 neutron& (0.5,+,+)  
   & --23.25   &  --25.41  &  --25.29  &  --25.68  & --25.11   &  --25.18   
   & --24.50   &  --24.90  &  --25.52  &  --24.92   \\
        & (0.5,--,+) 
   & --12.47   &   --8.72  &  --12.21  &   --9.17  & --11.31   &   --9.43   
   & --13.65   &  --14.76  &  --10.53  &  --14.76   \\
        & (1.5,+,--) 
   &  --5.20   &   --9.62  &   --7.18  &  --10.11  &  --8.58   &  --10.40   
   &  --7.64   &   --7.88  &  --11.69  &   --7.92   \\
        & (0.5,--,+) 
   &  --0.61   &   --2.34  &   --1.86  &   --2.70  &  --3.48   &   --3.67   
   &  --2.67   &   --3.07  &   --4.23  &   --3.11   \\
        & (0.5,+,+)  
   &  --1.54   &     0.81  &   --0.48  &     0.63  &    0.30   &     0.56   
   &  --2.15   &   --3.43  &     0.03  &   --3.44   \\
        & (0.5,--,+)  
   &    1.07   &     1.44  &     0.89  &     0.94  &    0.85   &     0.87   
   &    0.89   &     0.83  &     0.89  &     0.78   \\
        & (0.5,+,+)  
   &    1.04   &2.01(1.5+--)&    0.98  &     1.32  &    1.12   &     1.12   
   &    0.92   &     0.82  &     0.90  &     0.76   \\
        & (0.5,--,+)  
   & 1.46(0.5++)&    1.98  &     1.48   &     1.52   &1.37(0.5++)&    1.49  
   &1.03(1.5--\,--)&0.30(1.5--\,--)&0.45(2.5++)&0.27(1.5--\,--) \\
\hline 
\end{tabular} 
\end{table}   
\end{center}

The results are summarized in Table \ref{table1} and \ref{table2} for He isotopes
and in Table \ref{table3} and \ref{table4} for Be isotopes. 
The binding energies obtained here with Skyrme-IIIls interaction 
were overestimated than the corresponding empirical values 
\cite{alpha,rmfbec,bedata}.
However the CM energy correction to the binding energy for these nuclei are very 
large (more than one third of binding energy for He isotopes and more than one 
fifth for Be) and thus the binding energies have large uncertainty here due to 
the approximate treatment of CM energy.
Here the CM energy corrections were approximated by $1/A$ of total kinetic energy 
following the prescription used in Ref.\cite{he4xhf}. 
Approximating CM energy correction using harmonic oscillator energy \cite{cmho} 
makes the better agreement of binding energies with empirical values for Be isotopes
but it makes the binding energies too large for He isotopes.
Thus we examine here only the qualitative behavior of isotopes depending on the
neutron number. 
For a quantitative examination we first need to implement CM energy correction more 
self-consistently \cite{cmkin,cmfull} and need to search appropriate Skyrme interaction
which predicts correctly not only the binding energy but also other properties too
such as the size and deformation.

The binding energy and the root mean square (rms) radius of nuclear distribution 
show that $^{10}$Be is bound most strongly among Be isotopes considered here
while $^4$He is most stable among He isotopes.
Empirically, $^8$Be is bound stronger than $^{10}$Be.
However AMD-HF calculation \cite{beamdhf} also shows that $^{10}$Be is bound 
stronger than $^8$Be.
The rms radius of proton distribution is smallest for $^8$He among He isotope
and for $^{12}$Be among the ground states of Be isotopes.
Of course the rms radius of neutron distribution is larger for the isotope 
with more neutrons.
The values of the quadrupole moment show that the ground states of He and Be 
isotopes are prolate while the shape isomeric excited states of Be isotopes 
are oblate.
Even though we have not required the reflection symmetry in $z$-direction here,
the results show that the states of He and Be nuclei obtained here have a
reflection symmetry in the $z$-direction (i.e., $\pi_z$ is a good quantum number) 
and thus have a positive parity due to the filling of both $\pm\Omega$ levels of 
the even-even isotopes except for $^{10}$He and $^{18}$Be 
which have an asymmetric chain structure.

The single neutron levels in He and Be isotopes are somewhat insensitive to 
the number of neutrons in the isotope. 
Thus occupying the neutron levels from inner shell to outer shell makes 
the size of neutron distribution and the neutron skin larger for isotope 
with more neutrons.
In contrast to this behavior of neutrons in He and Be isotopes, the single 
proton levels are very sensitive to the number of neutrons.
The protons are bound stronger in the isotope with more neutrons up to $^8$He 
and up to $^{16}$Be.
This sensitivity of proton levels cannot be examined by any models assuming an 
inert core or $\alpha$ cluster.
AMD-HF calculations \cite{beamdhf} also show a similar behavior of the single 
particle energies of protons and neutrons in $^8$Be and $^{10}$Be.
In the oblate state of Be isotopes, the order of the single nucleon levels 
$(\Omega,\pi_z,\pi_r)=(1.5,+,-)$ and ($0.5,-,+$) of proton and of neutron is 
reversed from the order in the prolate state.
In a spherical limit, both of these two levels belong to the same $1p_{3/2}$ 
shell with different $\Omega$ values.
In the oblate excited state of Be isotopes, the level $(1.5,+,-)$ of proton and of 
neutron is lower than the level $(0.5,-,+)$ of proton and of neutron respectively 
and thus two protons occupy the level $(1.5,+,-)$ making the nuclei to be oblate. 
On the other hand, in the prolate ground state of He and Be isotopes, 
the level $(0.5,-,+)$ is lower than the level $(1.5,+,-)$ 
and thus two protons occupy the level $(0.5,-,+)$ making the nuclei to be prolate. 
This exhibits the existence of a level crossing when the nuclear shape changes from
the prolate state to the oblate state.

%%%   Figure 1 %%%
%
\begin{figure}
\caption{Contour plot of nuclear density distribution in He isotopes.
The densities for each contour curves are 0.15, 0.10, 0.05, 0.01, 0.005, 
0.001, 0.0005, and 0.00035 fm$^{-3}$ from inside of the contour.} \label{fig1}
\includegraphics[width=4.0in]{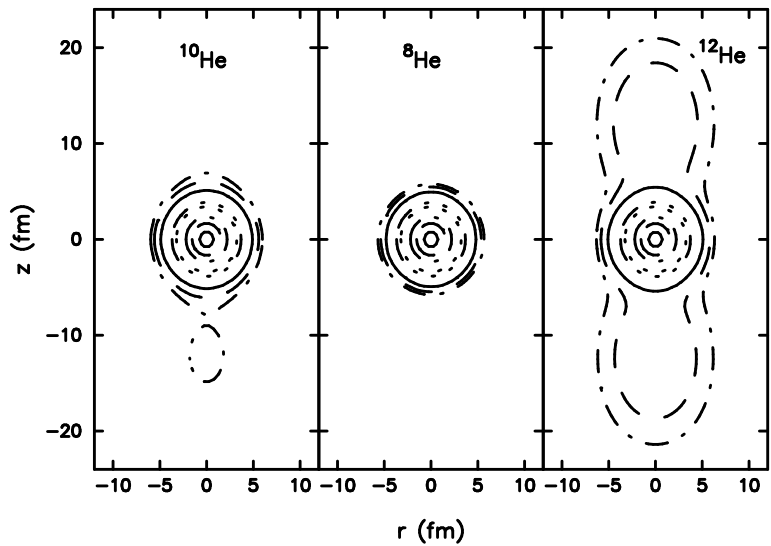}
\end{figure}

Even though it is insensitive to the number of neutrons, 
the rms radius of proton distribution shows that 
protons are most strongly bound in $^8$He among He isotopes considered here 
while the rms radius of neutron distribution increases as the number of 
neutrons increases. 
Proton distribution in He isotopes is spherical ($^4$He) or near spherical.
While the neutron distribution is spherical in $^4$He and near spherical in $^8$He,
it is deformed prolate in $^6$He.
The nuclear density profiles (see Fig.\ref{fig1}) show that the large rms radius 
and quadrupole moment of neutron distribution for $^{10}$He and $^{12}$He 
come from the $^8$He+2n and 2n+$^8$He+2n like chain structure respectively 
where the centers of each cluster are about 6 fm apart. 
Note that in Fig.\ref{fig1} for $^{10}$He we have a parity violating
intrinsic state consisting of a $^8$Be core and two neutrons mostly
on one side, i.e., showing an approximated $^8$He+2n like chain structure.
On the other hand $^{12}$He exhibits more exact 2n+$^8$He+2n like chain structure
with two 2n clusters located symmetrically. 
The neutrons in the correlated two neutron clusters are in the single nucleon 
levels with positive energy (see Table II) which means that they are unbound 
to the $^8$He cluster.
This fact is consistent with the empirical neutron separation energy \cite{bedata}; 
$^{10}$He has a negative two-neutron separation energy of --1.07 MeV  
and a small positive one-neutron separation energy of 0.2 MeV.
$^{10}$He and $^{12}$He may be either a resonance system or a Borromean system
\cite{borrome} (c.f., the one-neutron separation energy of $^9$He is --1.27 MeV).
Marqu\'{e}s et al. \cite{halosys} used two-neutron interferometry as a probe of
the neutron halo. However correlated neutrons can also be originated from a 
dineutron cluster or from a neutron skin which has mostly neutrons and is 
not necessarily a neutron halo. 

The single proton levels are very sensitive to the number of neutrons up to $^8$He
while the single neutron levels are somewhat insensitive.
The single nucleon levels in $^{10}$He and $^{12}$He stay similar to the levels 
in $^8$He.
The highest occupied neutron level is more strongly bound in $^8$He (2.28 MeV) 
than in $^6$He (1.96 MeV) 
and thus the neutron separation energy is larger for $^8$He than for $^6$He.
Empirically, one-neutron separation energies are 1.86 MeV and 2.57 MeV for $^6$He 
and $^8$He respectively while two-neutron separation energies are 0.97 MeV 
and 2.14 MeV respectively \cite{bedata,dripnuc}.
$^6$He has also a large neutron rms radius which is similar to the rms radius
of $^8$He.
This large neutron rms radius and small neutron separation energy
may indicate $^6$He is a halo nucleus.
However the large rms radius may come from a large deformation rather than
from a halo structure. The quadrupole moment of $^6$He is much larger than 
the quadrupole moments of $^4$He and $^8$He.
On the other hand, $^8$He, which is nearly spherical, has a larger 
neutron skin ($\Delta R = R_{\rm neutron} - R_{\rm proton} = 0.91$ fm)
than $^6$He and $^{16}$Be ground state 
and it has a larger nuclear rms radius than $^{12}$Be. 
The proton rms radius of $^8$He is the smallest among He isotopes and 
the nuclear quadrupole moment of $^8$He is smaller than of $^6$He. 
However $^8$He has a larger neutron separation energy than $^6$He
and the highest neutron level of $^8$He has angular momentum of at least $l = 1$
($\Omega = 1.5$ for this neutron level).
Thus more detailed analysis are required to check
if the ground state of $^8$He obtained here is actually a halo nucleus.

Be isotopes are prolate in their ground states and they have oblate shape 
excited states in which the nucleons filled self-consistently from 
the lowest single nucleon levels.
According to the binding energy per nucleon and the rms nuclear radius
$^{10}$Be is most strongly bound among the Be isotopes.
This suggests the shell closure at a neutron magic number 6 instead of 8. 
The vanishing of the neutron magic number 8 was predicted 
in Refs.\cite{ref10,magic6,fortun}
by studying the ground state and low-lying states of $^{12}$Be.
The appearance of a new magic number can be seen more directly
from the single particle levels.
There is a larger gap between the third ($1.5,+,-$) and the fourth ($0.5,-,+$) 
single neutron levels than the gap between the fourth ($0.5,-,+$) and 
the fifth (0.5,+,+) levels in He and Be isotopes (see Tables \ref{table2} 
and \ref{table4}) independent of which proton levels are occupied.
For a spherical or near spherical nuclei, this corresponds to a larger gap
between $1p_{3/2}$ and $1p_{1/2}$ shells than the gap between $1p_{1/2}$
and $2s_{1/2}$ or $1d_{5/2}$ shells.
It was suggested that the origin of the new magic number may be due to neutron
halo formation \cite{magicprl}. But our calculation shows that the new magic number
originates from a nuclear deformation.
The gap between the fourth level ($0.5,-,+$) and the fifth level ($0.5,+,+$) is 
smaller for an isotope with a larger deformation, especially, of proton distribution.
For a largely deformed nuclei, $^8$Be and $^{16}$Be, a level inversion between
the fifth and the fourth single particle levels occurs.
Thus the intruder state $1/2^+$ causing the inversion between $1/2^+$ and $1/2^-$ 
for the ground state of $^{11}$Be and $^9$He \cite{be10excit,magic16}
may be originated from a large deformation of proton distribution
independent of which levels of proton and neutron are occupied.
It was also shown \cite{intrudbe} that low-lying intruder states of $^8$Be, $^{10}$Be,
and $^{12}$C can be described better with a deformed oscillator model than
with a spherical shell model.

The ground state of $^{12}$Be has the smallest proton rms radius and 
the most spherical nuclear distribution among Be isotopes in its ground state. 
The density distribution shows that $^{12}$Be has a small peanut shell 
shaped core of higher central density surrounded by
a large spherical neutron skin 
of lower density without showing any increase of neutron density in radius.
The proton levels $(\Omega, \pi_z, \pi_r) = (0.5, -, +)$ and $(1.5,+,-)$ 
in the ground state of $^{12}$Be are nearly degenerated 
and thus the proton shell is unclosed 
in this calculation with Skyrme-IIIls interaction. 
In the spherical limit, both of these two levels belong to the $1p_{3/2}$ level and 
thus are exactly degenerated. 
This degeneracy causes the excited oblate state to be nearly degenerated with
the prolate ground state for $^{12}$Be (see Table \ref{table3} and discussion below). 
The ground state of $^8$Be has a large quadrupole moment with a peanut shell (with
two kernels) shaped nuclear distribution which may be considered as
a two-$\alpha$ chain structure.
The nuclear shape of $^{10}$Be and $^{14}$Be are deformed prolate but $^{10}$Be has
a small peanut shell shaped core.
$^{16}$Be has a peanut shell shaped nuclear distribution and $^{18}$Be has 
an approximated $^{16}$Be+2n like chain structure. 
The neutron rms radius of Be isotopes increases as the number of neutrons increases.
The large neutron rms radius and quadrupole moment of $^{18}$Be come from 
the $^{16}$Be+2n like chain structure. 
The single nucleon levels in $^{18}$Be stay similar to the levels in $^{16}$Be.
Two neutrons occupying the highest single particle level ($1.5,-,-$) in $^{16}$Be 
have a small positive energy but without showing any increase of neutron density 
in the large neutron skin.
Thus, $^{16}$Be is beyond the neutron drip line while $^{14}$Be is inside of 
the drip line in these calculations. 
Jensen and Riisager \cite{neutclus} pointed out the possible existence
of exotic nuclei even beyond the neutron drip line due to a correlation 
between many neutrons.
Empirically, $^{16}$Be has a negative two-neutron separation energy of --1.58 MeV  
and a small positive one-neutron separation energy of 0.19 MeV \cite{bedata}.
In $^{18}$Be four neutrons occupy single neutron levels with positive energy 
in the nucleus.
The ground states of $^{16}$Be and $^{18}$Be with neutrons in positive energy level
may be either one of Borromean nucleus \cite{borrome} or one of resonance.
To check this, more detailed investigations are needed.

Be isotopes in their excited states have smaller rms radii and smaller 
magnitude of quadrupole moments than in their ground states.
A four-body cluster model based on a resonating group model \cite{be10excit} 
also predicted the $2_1^+$ excited state of $^{10}$Be has smaller rms radius 
than the $0^+$ ground state.
The excitation energies are 5.4, 2.1, 0.1, and 3.8 MeV 
for $^8$Be, $^{10}$Be, $^{12}$Be, and $^{16}$Be respectively.
However, since we worked with non-spherical quantum numbers with axial symmetry,
we cannot directly compare the excited states obtained here for Be isotopes 
with experimental results which are usually identified by spherical quantum numbers.
The nearly degenerated (0.1 MeV excitation) oblate and prolate states of $^{12}$Be
may cause a shape coexistence \cite{cmfull} for the ground state of $^{12}$Be.

The lowest single nucleon level in each isotope is more strongly bound in the oblate 
excited state than in the prolate ground state while the next nucleon level 
is higher in the oblate shape isomeric state than in the prolate ground state.
The third nucleon level of each isotope is also more strongly bound in the 
oblate excited state than in the prolate ground state. 
Thus the energy gap between the second level $(\Omega, \pi_z, \pi_r) = (0.5, -, +)$ 
and the third level $(1.5,+,-)$ of proton and neutron in Be isotopes
is larger in their prolate ground state (more than 2.5 MeV) 
than in their oblate shape isomeric state (less than about 1 MeV)
except for the proton levels in $^{12}$Be.  
This result is consistent with the fact that these two levels belong to a
spherical $1p_{3/2}$ shell and the Be isotopes in their oblate excited states 
are more spherical than in their prolate ground states.
In $^{12}$Be with Skyrme-IIIls interaction, these two proton levels are nearly 
degenerated with the gap of 0.02 MeV in the prolate ground state
while the gap is 0.88 MeV in the oblate excited state.
The order of these single nucleon levels $(0.5, -, +)$ and $(1.5,+, -)$ 
in the excited oblate state of Be isotopes are reversed from the order in the
prolate ground state except for proton levels in $^{12}$Be.
In the excited oblate state of $^{12}$Be, the unoccupied $(0.5,-,+)$ 
proton level is lower than the occupied $(1.5,+,-)$ proton level.
The protons in the excited oblate state of $^{12}$Be is not filled self-consistently 
from the lowest single proton levels 
but is filled in a fixed configuration
in this calculation with Skyrme-IIIls interaction. 
Thus the excited oblate state of $^{12}$Be obtained here is a proton $2p-2h$ 
excited state instead of a shape isomeric state.
All other states of He and Be isotopes in these
calculations are formed by filling protons and neutrons self-consistently from the
lowest single nucleon levels. 

While the same neutron levels of Be isotopes are occupied both in the prolate 
ground state and in the oblate excited state except for $^8$Be,
different proton levels are occupied in the ground state and in the excited state.
The Be isotopes become prolate by filling the $(0.5,-,+)$ proton level and
become oblate by filling the $(1.5,+,-)$ proton level.
$^{12}$Be is the most spherical among Be isotopes 
both in their ground states and in their excited states.
The excited state of $^{16}$Be has a large neutron rms radius (3.97 fm) 
and a large neutron skin ($\Delta R = 1.56$ fm) with relatively small
quadrupole moment and four neutrons in positive energy levels.
This probably indicates the excited oblate $^{16}$Be is a halo nucleus.
However more detailed analysis are required to check
if the excited oblate state of $^{16}$Be obtained here is actually a halo nucleus.

In conclusion of this paper 
we have investigated the mean field structure of He and Be isotopes 
in a nonrelativistic Hartree-Fock approximation with Skyrme-IIIls interaction
in an axially symmetric configuration space.
The ground and shape isomeric states of these isotopes are obtained by occupying 
protons and neutrons self-consistently from the lowest single nucleon levels.
It is shown that the ground states of He and Be isotopes are prolate 
while the excited shape isomeric states of Be isotopes are oblate. 
There is a level crossing between the second and the third nucleon levels, 
$(\Omega,\pi_z,\pi_r) = (0.5,-,+)$ and $(1.5,+,-)$  
which belong to a spherical $1p_{3/2}$ shell, 
when the isotopes change from the prolate ground state to the oblate excited state.
The level $(0.5,-,+)$ of proton and of neutron is lower than the level $(1.5,+,-)$ 
in the prolate ground states and the level $(1.5,+,-)$ is lower 
in the oblate excited states of Be isotopes.
Thus two protons occupy the level $(0.5,-,+)$ in the prolate 
ground states making the nuclei prolate while they occupy the level $(1.5,+,-)$ 
in the oblate excited states making the nuclei oblate.
Also shown in these simple calculations is that the protons are bound stronger 
in the isotope with more neutrons while neutron levels are somewhat insensitive 
to the number of neutrons and thus the nuclear size and also the neutron skin 
become larger as the neutron number increases. 
The binding energy of Be isotopes and the single neutron levels obtained here
indicate the existence of the neutron magic number 6 instead of 8. 
The vanishing of the neutron magic number 8 was predicted in Refs.\cite{ref10,magic6,fortun}.
The behavior of the gap between the fourth and fifth neutron levels obtained here
suggests that the disappearance of the magic number 8 is originated 
from a large deformation of nucleus.
It is also shown that a level inversion between the fourth ($0.5,-,+$) and 
the fifth ($0.5,+,+$) levels occurs only in a largely deformed isotope, 
i.e., in $^8$Be and $^{16}$Be.
This inversion is related to the intruder state $1/2^+$ in the
ground state of $^{11}$Be and $^9$He \cite{be10excit,magic16}.
In these calculations for He and Be isotopes with Skyrme-IIIls interaction, 
no system with a clear indication of neutron halo was found. 
Instead of it we found $^8$He+2n, 2n+$^8$He+2n, and $^{16}$Be+2n 
like chain structures with clusters of two correlated neutrons 
for $^{10}$He, $^{12}$He, and $^{18}$Be respectively.
The correlated neutrons in the dineutron cluster have a positive single particle 
energy and are not bound to the $^8$He or $^{16}$Be cluster.
These nuclei with chain structure may be either a resonance or one of Borromean 
nucleus \cite{borrome}. More detailed investigations are needed.
It is also shown that the nuclei $^8$He and $^{14}$Be are below the neutron drip line
in which all nucleons are bound (with negative energy).
We have also found that $^{16}$Be in its ground state, 
as a single cluster without 2n correlated cluster, 
is beyond the neutron drip line with two neutrons in positive energy levels. 
Due to two neutrons in a positive energy level, $^{10}$He 
and $^{16}$Be have a negative two-neutron separation energy
in agreement with data \cite{bedata}. 
It is also shown that the neutron separation energy of $^6$He is smaller
than the separation energy of $^8$He which is consistent with data \cite{bedata}.
The large rms radius of neutron distribution in $^6$He, which is similar to 
the rms radius of $^8$He, is originated from the large deformation of $^6$He
rather than from a halo structure.
The most probable candidate of halo nucleus obtained here is 
the excited oblate state of $^{16}$He which has a large neutron skin of 1.56 fm 
with a relatively small quadrupole moment.

We should notice here that the 
calculations done here cannot examine a nuclear structure 
with multi-centered clusters such as $^6$He+$^6$He or $^8$He+$^4$He. 
The $2\alpha$ cluster structure in $^{12}$Be was found in Refs.\cite{ref8,ref9,ref10}
even though they missed the detailed nuclear structure.
It was also shown, using relativistic mean field theory, that an excited state of $^{12}$C 
is a chain structure of $^8$Be and $\alpha$ clusters centered at two points 
while the ground state is oblate with a single center \cite{alpha}.
To study various states of these nuclei properly in a mean field approach, 
we need to extend our HF calculation to allow 
nuclear structures with multi-centered clusters \cite{alpha}.
On the other hand 
to study halo structure we need to extend our calculation to allow highly excited
single nucleon levels and to consider other nuclei such as C and O isotopes too.
Notice here also that the binding energies obtained here with the Skyrme-IIIls 
interaction were overestimated with a large uncertainty
due an approximate treatment of CM energy correction.
But we believe that the main features of isotopes obtained here would not
be altered when we use other mean field potentials with a proper
treatment of CM energy correction. 
However, for a quantitative study in future, we need to implement CM energy 
correction more exactly and also need to investigate the dependence of 
the various behaviors of isotopes on the mean field potential used.

This work was supported 
by Grant No. KHU-20050313 of the Kyung Hee University Research Fund in 2005.
The authors thank L. Zamick for helpful comments and discussions.

\end{document}